\documentclass[twocolumn,aps,pra,longbibliography,superscriptaddress]{revtex4-2}
\usepackage{amsmath,latexsym}
\usepackage{xcolor}
\usepackage[%
  colorlinks=true,
  urlcolor=blue,
  linkcolor=blue,
  citecolor=blue
]{hyperref}
\usepackage{etoolbox}
\usepackage{breqn}
\usepackage{mathrsfs,verbatim,mathtools,graphicx,amsmath, amssymb, bm, epstopdf, enumerate}
\makeatletter
\let\cat@comma@active\@empty
\makeatother

\begin{document}

\title{Efficient characterization of spatial Schmidt modes of multiphoton entangled states produced from high-gain parametric down-conversion}

\author{Mahtab Amooei}
\email{mamoo074@uottawa.ca}
\affiliation{Department of Physics, University of Ottawa, Ottawa, Ontario K1N 6N5, Canada}
\author{Girish Kulkarni}
\email{Present address: Department of Physics, Indian Institute of Technology Ropar, Rupnagar - 140001, Punjab, India}
\affiliation{Department of Physics, University of Ottawa, Ottawa, Ontario K1N 6N5, Canada}
\author{Jeremy Upham}
\affiliation{Department of Physics, University of Ottawa, Ottawa, Ontario K1N 6N5, Canada}
\author{Robert W. Boyd}
\affiliation{Department of Physics, University of Ottawa, Ottawa, Ontario K1N 6N5, Canada}
\affiliation{The Institute of Optics, University of Rochester, Rochester, New York 14627, USA}
\date{\today}
\begin{abstract}
The ability to efficiently characterize the spatial correlations of entangled states of light is critical for applications of many quantum technologies such as quantum imaging. Here, we demonstrate highly efficient theoretical and experimental characterization of the spatial Schmidt modes and the Schmidt spectrum of bright multiphoton entangled states of light produced from high-gain parametric down-conversion. In contrast to previous studies, we exploit the approximate quasi-homogeneity and isotropy of the signal field and dramatically reduce the numerical computations involved in the experimental and theoretical characterization procedures. In our particular case where our experimental data sets consist of 5000 single-shot images of 256$\times$256 pixels each, our method reduced the overall computation time by 2 orders of magnitude. This speed-up would be even more dramatic for larger input sizes. Consequently, we are able to rapidly characterize the Schmidt modes and Schmidt spectrum for a range of pump amplitudes and study their variation with increasing gain. Our results clearly reveal the broadening of the Schmidt modes and narrowing of the Schmidt spectrum for increasing gain with good agreement between theory and experiment.
\end{abstract}
\maketitle
\section{Introduction}
One of the major thrusts in the field of quantum optics is the efficient generation and characterization of entangled and squeezed states of light for a variety of quantum applications such as communication \cite{bouwmeester1997nature, furusawa1998science}, computation \cite{carolan2015science, madsen2022nature}, sensing \cite{pirandola2018natphot,lawrie2019acsphot}, imaging \cite{moreau2019natrevphys}, and phase metrology \cite{caves1981prd, mcculler2020prl}. In this context, the most widely used experimental platform is a process known as spontaneous parametric down-conversion (SPDC) in which an incident pump field interacts with a nonlinear optical medium to produce a pair of fields known as the signal and idler \cite{klyshko1967jetp,burnham1970prl}. For low pump strengths, the output signal-idler quantum state is approximately an entangled two-photon state \cite{hong1985pra}, whereas for high pump strengths, it is a multiphoton entangled bipartite state known as bright squeezed vacuum \cite{agafonov2010pra}. In both these regimes of SPDC, the spatial and photon number correlations between the signal and idler fields have been of particular interest for quantum imaging schemes such as sub-shot noise quantum imaging \cite{brida2010natphot}, ghost imaging \cite{pittman1995pra}, and imaging with undetected photons \cite{lemos2014nature}. In such schemes, the critical performance parameters such as imaging resolution and contrast ultimately depend on the spatial correlations of the entangled fields \cite{ruo2020apl, moreau2018opex, fuenzalida2022quantum}. Therefore, in order to be able to devise strategies for optimizing these parameters while simultaneously preserving any inherent quantum advantage, it is vital to characterize these spatial correlations and their variation with increasing pump strength.

An important tool used for characterizing the spatial correlations of entangled pure states produced from SPDC is the Schmidt decomposition \cite{miatto2012epjd1}. This decomposition uniquely expresses every bipartite state in terms of its constituent Schmidt modes that are weighted by the Schmidt spectrum, whose effective width quantifies the degree of entanglement. In the context of low-gain SPDC, previous studies have analytically shown that when the sinc-shaped phase-matching function is approximated by a Gaussian function, the Schmidt modes of the spatial two-photon wavefunction are the Laguerre-Gaussian functions in polar coordinates, or equivalently, the Hermite-Gaussian functions in Cartesian coordinates \cite{law2004prl,exter2006pra,miatto2012epjd2,straupe2011pra}. The polar representation has been the focus of studies on the orbital angular momentum (OAM) spectrum and its dependence on pump and crystal parameters \cite{pires2010prl,kulkarni2017natcomm,kulkarni2018pra}, but the Cartesian representation has also been studied to some extent \cite{exter2006pra,miatto2012epjd2,straupe2011pra}.

In contrast to low-gain SPDC, the theoretical  characterization of Schmidt modes in high-gain SPDC has been challenging because no closed-form analytic solution of the output state exists at present for a general pump field \cite{brambilla2004pra, sharapova2015pra, sharapova2020prr, kulkarni2022prr}. In a previous study, a model of high-gain SPDC was developed by assuming that the size and shape of the Schmidt modes are independent of the gain \cite{sharapova2015pra}. However, the model was unable to explain the experimentally observed broadening of the far-field intensity profile with increasing gain indicating that the underlying assumption was invalid. Subsequently, Ref.~\cite{sharapova2020prr} relaxed the assumption and carried out a numerical calculation of 1D slices of the Schmidt modes and showed their profiles to be in agreement with corresponding experimentally inferred 1D slices through intensity covariance measurements. These measurements revealed a slight expansion of the individual Schmidt modes with increasing gain.

Recently, one study has performed the experimental reconstruction of the full 2D Schmidt modes for a fixed pump strength in the high-gain regime \cite{averchenko2020pra}. In this reconstruction method, one first acquires a large number of $(N\times N)$-pixeled images of the signal in the far-field, computes the $(N\times N\times N\times N)$ 4D spatial correlation function from intensity covariance measurements, and then diagonalizes this correlation function to yield the $(N\times N)$-pixeled 2D Schmidt eigenmodes. As the spatial resolution and the spectral range of the measurement both increase with $N$, it is desirable to maximize $N$ for a given transverse size of the measured field. However, the computation and diagonalization of the 4D spatial correlation function become highly cumbersome and impractical as $N$ increases. Therefore, there is a strong need to reduce the computational complexity, at least for specific experimental scenarios that are commonly encountered.

In this article, we consider the common scenario involving a rotationally symmetric pump field undergoing SPDC to produce a quasi-homogeneous and isotropic signal field. In particular, we assume that the normalized degree of spatial coherence of the signal field is translationally and rotationally invariant. In this case, we show that the translational and rotational symmetries can be exploited to vastly enhance the computational efficiency of the aforementioned characterization method. In particular, instead of computing the full $(N\times N\times N\times N)$ 4D spatial correlation function, we now only compute an $(N\times N)$ 2D spatial correlation function, diagonalize it to obtain $(N\times 1)$ 1D Schmidt modes, and then tensor-multiply those modes to obtain the full $(N \times N)$ 2D Schmidt modes. As a result, we are able to characterize the spatial Schmidt modes both theoretically and experimentally for a range of increasing pump strengths in a highly efficient manner. The paper is organized as follows: In Sec.~\ref{theory} we present a theoretical description of the characterization procedure. In Sec.~\ref{experiment}, we present our experimental results and their comparison with theoretical predictions. In Sec.~\ref{conclusions}, we conclude with a summary and outlook of our study.
\section{Theoretical Analysis}\label{theory}
We consider the case of near-degenerate type-I SPDC from a quasi-monochromatic pump field. In this situation, we can suppress the frequency and polarization variables and write down the Schmidt decomposition of the joint spatial bipartite wavefunction $\Psi(\bm{q}_{s},\bm{q}_{i})$ of the signal and idler fields as
\begin{align}\label{schmidt-decomp}
\Psi(\bm{q}_{s},\bm{q}_{i})=\sum_{k=0}^{\infty}\sqrt{\lambda_{k}}\,\bm{u}_{k}(\bm{q}_{s})\bm{u}_{k}(\bm{q}_{i}),
\end{align}
where the positive real numbers $\lambda_{k}$ constitute the Schmidt spectrum, $\bm{q}_{j}$ represents the transverse wavevector, and the normalized spatial functions $\bm{u}_{k}(\bm{q}_{j})$ constitute the Schmidt modes for $j=s,i$ corresponding to the signal and idler fields, respectively. As the two fields have been assumed to be nearly degenerate in frequency and identically polarized, the situation is nearly symmetric under an exchange of the signal and idler fields, and therefore, the Schmidt modes $\bm{u}_{k}$ for both the fields can be regarded to be identical. In the low-gain regime, the modes are populated by one photon each, whereas in the high-gain regime, the same modes can be populated by higher but equal numbers of photons. Both the Schmidt spectrum $\left\{\lambda_{k}\right\}$ and the Schmidt modes $\left\{\bm{u}_{k}\right\}$, depend on the gain of the process.
\begin{figure}[t]
	\includegraphics[scale=0.5]{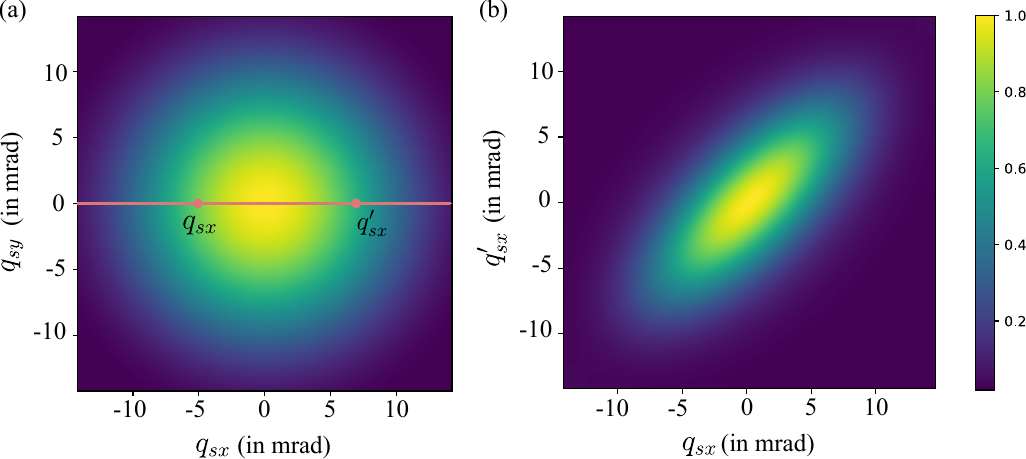}
	\caption{Numerically simulations of SPDC of a Gaussian pump for a typical gain value in high-gain limit of parametric down-conversion: (a) far-field intensity profile $I(\bm{q}_{s})=G^{(1)}(\bm{q}_{s},\bm{q}_{s})$, and (b) Real part of the first-order spatial correlation function $G^{(1)}(q_{sx},q'_{sx})$.}
    \label{fig1}
\end{figure}
In terms of the above definitions, the first-order spatial correlation function $G^{(1)}(\bm{q}_{s},\bm{q'}_{s})$ of the signal field can be shown to take the form \cite{glauber1963pr, orderconv}
\begin{align}\label{coherent-mode-decomp}
G^{(1)}(\bm{q}_{s},\bm{q'}_{s})=\sum_{k=0}^{\infty}\lambda_{k}\,\bm{u}_{k}(\bm{q}_{s})\,\bm{u}^*_{k}(\bm{q}_{s}'),
\end{align}
which is formally equivalent to the reduced density matrix of the signal field obtained by performing a partial trace of the global bipartite pure state over the idler modes. The above equation expresses the coherent mode decomposition of the signal field. Thus, the Schmidt modes and the Schmidt spectrum of the entangled signal-idler state from Eq.~(\ref{schmidt-decomp}) can be directly obtained as the eigenmodes and eigenvalues of the coherent mode decomposition Eq.~(\ref{coherent-mode-decomp}) of the signal field, respectively. This connection between the Schmidt decomposition of the global bipartite pure state and the coherent mode decomposition of the individual fields has been extensively used in previous studies for characterizing the entanglement of bipartite pure states \cite{pires2009pra,jha2011pra,kulkarni2017natcomm,kulkarni2018pra,bhattacharjee2022njp}.
\begin{figure*}[t!]
\centering
	\includegraphics[width=\textwidth]{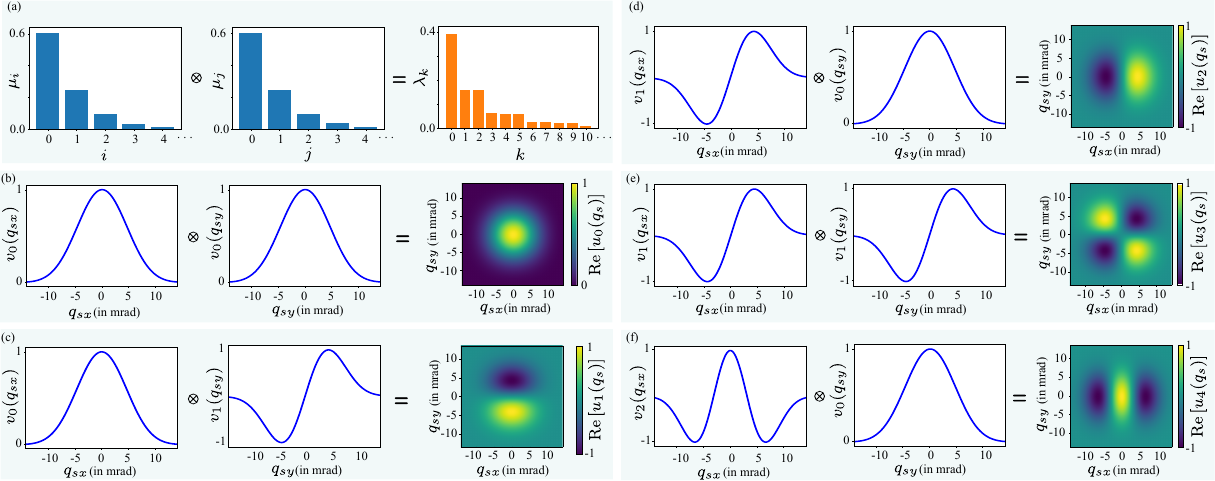}
	\caption{Illustration of the procedure for obtaining (a) the Schmidt spectrum, and (b)-(f) 2D Schmidt modes through a tensor product of their 1D counterparts for the first few modes.}
    \label{fig2}
\end{figure*}

The theoretical form of the first-order spatial correlation function of the signal field produced from high-gain SPDC is given by \cite{kulkarni2022prr}
\begin{align}\notag
G^{(1)}(\bm{q}_{s},\bm{q'}_{s})&=\frac{C_{1}}{k_{sz}k'_{sz}}\int\mathrm{d}{\bm \rho}\,\langle|V_{p}({\bm \rho})|^2\rangle\,e^{-i[({\bm q_{s}}-{\bm q'_{s}})\cdot{\bm \rho}]}\\\notag&\hspace{-1cm}\times\left[\frac{\mathrm{sinh}\,\Gamma(\Delta \bar{k}_{z},{\bm \rho})L}{\Gamma(\Delta \bar{k}_{z},{\bm \rho})}\right]\left[\frac{\mathrm{sinh}\,\Gamma(\Delta \bar{k'}_{z},{\bm \rho})L}{\Gamma(\Delta \bar{k'}_{z},{\bm \rho})}\right]\\\label{hgpdccorrfunc}&\hspace{2cm}\times e^{i(\Delta \bar{k}_{z}-\Delta \bar{k}'_{z})L/2},
\end{align}
where $k_{sz}$ is the longitudinal wavevector component, ${\bm \rho}$ is the transverse position vector at the crystal plane, $V_{p}({\bm \rho})$ is the pump transverse amplitude profile, $L$ is the crystal length, $\Delta \bar{k}_{z}$ is the longitudinal wavevector mismatch $\Delta k_{z}$ evaluated under the condition ${\bm q_{s}}+{\bm q_{i}}=0$ as discussed in Ref.~ \cite{kulkarni2022prr}, and $C_{1}$ is an overall constant. The quantity $\Gamma(\Delta \bar{k}_{z},{\bm \rho})$ takes the form \cite{kulkarni2022prr}
\begin{align}\label{gammaexp}
\Gamma(\Delta \bar{k}_{z},{\bm \rho})\equiv\left[\frac{C_{2}}{k_{sz}\bar{k}_{iz}}|V_{p}({\bm \rho})|^2-\left(\frac{\Delta \bar{k}_{z}}{2}\right)^2\right]^{1/2},
\end{align}
where $\bar{k}_{iz}$ is the longitudinal wavevector component $k_{iz}$ of the idler field evaluated under the condition ${\bm q_{s}}+{\bm q_{i}}=0$, and $C_{2}$ is a scaling factor. Thus, using equations (\ref{hgpdccorrfunc}) and (\ref{gammaexp}), it is possible to compute $G^{(1)}(\bm{q}_{s},\bm{q'}_{s})$ for a pump field with any arbitrary strength and spatial profile.

We now note that, in general, the function $G^{(1)}(\bm{q}_{s},\bm{q'}_{s})$ needs to be evaluated for all pairs of transverse wavevectors $\bm{q}_{s}$ and $\bm{q'}_{s}$ with each evaluation involving an integration over the spatial position vector ${\bm \rho}$, which is a cumbersome procedure. Furthermore, the subsequent diagonalization of $G^{(1)}(\bm{q}_{s},\bm{q'}_{s})$ to yield the Schmidt spectrum and Schmidt modes can also be highly computationally intensive \cite{averchenko2020pra}. However, let us consider the special case of a rotationally symmetric pump field, i.e, $V_{p}({\bm \rho})=V_{p}(\rho)$. If the crystal is sufficiently thin such that the transverse walk-off is negligible, the generated field can be approximated to be quasi-homogeneous and isotropic, which leads to the condition \cite{pires2009pra}
\begin{align}\label{g1degcoh}
 G^{(1)}(\bm{q}_{s},\bm{q'}_{s})\approx\sqrt{I(\bm{q}_{s})I(\bm{q'}_{s})}\,g^{(1)}(|\bm{q}_{s}-\bm{q'}_{s}|),
\end{align}
where $I(\bm{q}_{s})= G^{(1)}(\bm{q}_{s},\bm{q}_{s})$ is the far-field intensity as depicted in Fig.~\ref{fig1}(a) and $g^{(1)}(|\bm{q}_{s}-\bm{q'}_{s}|)$ is the degree of spatial coherence. In other words, the degree of spatial coherence $g^{(1)}$ depends only on the relative distance $|\bm{q}_{s}-\bm{q'}_{s}|$ between $\bm{q}_{s}$ and $\bm{q'}_{s}$, and not on either of them individually. As a result, it is sufficient to compute $G^{(1)}(q_{sx},q'_{sx})$ as depicted in Fig.~\ref{fig1}(b) along just a single axis, say the x-axis, to infer the statistics of the field, which is a much simpler task than computing the entire function $G^{(1)}(\bm{q}_{s},\bm{q}_{s})$. Furthermore, the 2D function $G^{(1)}(q_{sx},q'_{sx})$ can be readily diagonalized as
\begin{align}
 G^{(1)}(q_{sx},q'_{sx})=\sum_{i=0}^{\infty}\mu_{i}\,v_{i}(q_{sx})v_{i}^*(q'_{sx}),
\end{align}
where $\mu_{i}$ are positive real coefficients and $v_{i}(q_{sx})$ are 1D normalized eigenfunctions. From rotational invariance, it follows that the corresponding eigenvalues and eigenfunctions of the y-axis counterpart $G^{(1)}(q_{sy},q'_{sy})$ would be identical to those for the x-axis. The full Schmidt spectrum $\left\{\lambda_{k}\right\}$ and the full Schmidt modes $\left\{\bm{u}_{k}\right\}$ Schmidt spectrum can then be computed as
\begin{subequations}
 \begin{align}
  \left\{\lambda_{k}\right\}&=\left\{\mu_{i}\,\mu_{j}\right\},\\
  \left\{\bm{u}_{k}(\bm{q}_{s})\right\}&=\left\{v_{i}(q_{sx})\otimes v_{j}(q_{sy})\right\}.
 \end{align}
\end{subequations}
In other words, the Schmidt spectrum $\left\{\lambda_{k}\right\}$ is computed by taking the tensor product of the $\left\{\mu_{i}\right\}$ with itself and then arranging the elements in non-increasing order. Similarly, the full 2D Schmidt modes can be computed by taking a pair-wise tensor product of the 1D Schmidt modes with themselves.
\begin{figure*}[t!]
\centering
	\includegraphics[width=0.96\textwidth]{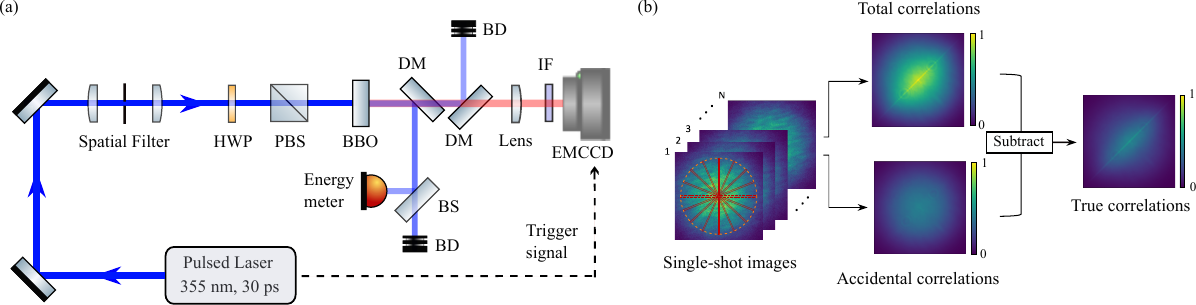}
	\caption{(a) Schematic of the experimental setup, (b) Conceptual depiction of the procedure for experimentally measuring $G^{(1)}(q_{sx},q'_{sx})$ from multiple diametric slices of multiple single-shot images of the signal field. HWP: half-wave plate, PBS: polarizing beam-splitter, BBO: $\beta$-barium borate, DM: dichroic mirror, BD: beam dump, IF: interference filter of bandwidth $10$ nm centered at $700$ nm, EMCCD: electron-multiplying charge coupled device camera.}
    \label{fig3}
\end{figure*}

For concreteness, we consider a specific case, namely that of type-I collinear SPDC from a Gaussian pump field. We carry out numerical simulations for the following parameters:
\begin{subequations}
\begin{align}
 V_{p}({\bm \rho})&=g\exp\{-\rho^2/w^2_{p}\},\\
\Delta \bar{k}_{z}&=|{\bm k_{p}}|-2\sqrt{|{\bm k_{s}}|^2 -|{\bm q_{s}}|^2},
\end{align}
\end{subequations}
where $g$ is the pump amplitude and $w_{p}=185\,\mu$m is the $1/e^2$ pump spatial beam-waist size. We assume pump wavelength $\lambda_{p}=355$ nm, signal wavelength $\lambda_{s}=700$ nm, and $L=3$ mm is the length of the crystal, which we assume to be of $\beta$-barium borate (BBO). The quantities $|{\bm k_{s}}|=2\pi n_{so}/\lambda_{s}$ and $|{\bm k_{p}}|=2\pi\eta_{p}(\theta_{p})/\lambda_{p}$ are the signal and pump wavevector magnitudes, respectively. Here, $\theta_{p}$ is the angle between the pump propagation direction inside the crystal and its optic axis, and $\eta_{p}(\theta_{p})=n_{pe}n_{po}/\sqrt{n_{po}^2\sin^2\theta_{p}+n_{pe}^2\cos^2\theta_{p}}$ is the effective refractive index of the extraordinary-polarized pump inside the crystal \cite{walborn2010physrep}. The values $n_{p(e)o}$ and $n_{so}$ of the (extra)ordinary and ordinary refractive indices of BBO for the pump and signal wavelengths, respectively, can be obtained using the Sellmeier relations \cite{eimerl1987jap}
\begin{subequations}\label{bbodisrelations}
 \begin{align}
  n_{e}^2(\lambda)&=2.7405+\frac{0.0184}{\lambda^2-0.0179}-0.0155\lambda^2,\\
  n_{o}^2(\lambda)&=2.3730+\frac{0.0128}{\lambda^2-0.0156}-0.0044\lambda^2,
 \end{align}
\end{subequations}
where $\lambda$ is the corresponding wavelength in micrometers. We substitute these relations in Eq.~(\ref{hgpdccorrfunc}) to compute the intensity profile $I({\bm q_{s}})$ and $G^{(1)}(q_{sx},q'_{sx})$, which we depict in Figs.~\ref{fig1}(a) and \ref{fig1}(b), respectively. We then diagonalize $G^{(1)}(q_{sx},q'_{sx})$ and tensor-multiply the eigenvalues and eigenmodes to compute the full Schmidt spectrum and 2D Schmidt modes as depicted in Fig.~\ref{fig2}. Note that there are double degeneracies observed in the distribution $\left\{\lambda_{k}\right\}$ due to the product $\mu_{i}\,\mu_{j}$ being the same for the two possible permutations of $\mu_{i}$ and $\mu_{j}$. Moreover, the corresponding Schmidt modes $v_{i}(q_{sx})\otimes v_{j}(q_{sy})$ and $v_{j}(q_{sx})\otimes v_{i}(q_{sy})$ are also equivalent up to a rotation of $90^{\circ}$. These degeneracies and permutation equivalences originate from our assumption of rotational invariance.

\section{Experimental results}\label{experiment}
We now consider the task of performing the experimental characterization of the Schmidt spectrum and Schmidt modes. In general, an experimental measurement of $G^{(1)}(\bm{q}_{s},\bm{q'}_{s})$ necessarily involves the use of interferometry. However, for our purposes, this is not essential. As the global state of the signal-idler field is a bipartite squeezed pure state, it follows that the individual state of the signal field obeys thermal statistics \cite{loudon2000book}. For a thermal field, there is a simple relation between $G^{(1)}(\bm{q}_{s},\bm{q'}_{s})$ and the second-order spatial correlation function $G^{(2)}(\bm{q}_{s},\bm{q'}_{s})\equiv\langle I(\bm{q}_{s})I(\bm{q'}_{s})\rangle$, where $\langle \cdots\rangle$ denotes an ensemble average over many realizations of the field. This relation is known as the Siegert relation and can be written as (see Sec. 3.7 of \cite{loudon2000book})
\begin{align}\label{siegert}
 G^{(2)}(\bm{q}_{s},\bm{q'}_{s})=|G^{(1)}(\bm{q}_{s},\bm{q'}_{s})|^2\,+\,\langle I(\bm{q}_{s})\rangle \langle I(\bm{q'}_{s})\rangle.
\end{align}
Thus, for a thermal field, the magnitude of $G^{(1)}(\bm{q}_{s},\bm{q'}_{s})$ can be inferred from a characterization of $G^{(2)}(\bm{q}_{s},\bm{q'}_{s})$ using intensity correlation measurements. For a rotationally symmetric source, the Fourier relationship between the near-field and far-field correlations imply that $G^{(1)}(\bm{q}_{s},\bm{q'}_{s})$ can be approximated to be real, i.e, $G^{(1)}(\bm{q}_{s},\bm{q'}_{s})\approx |G^{(1)}(\bm{q}_{s},\bm{q'}_{s})|$ \cite{mandel1995cup,pires2009pra2,averchenko2020pra}. We also verified the validity of this approximation by numerically computing $G^{(1)}(\bm{q}_{s},\bm{q'}_{s})$ using Eq.~(\ref{hgpdccorrfunc}) for a Gaussian pump field. Substituting $G^{(2)}(\bm{q}_{s},\bm{q'}_{s})\equiv\langle I(\bm{q}_{s})I(\bm{q'}_{s})\rangle$ and $\delta I(\bm{q}_{s})\equiv I(\bm{q}_{s})-\langle I(\bm{q}_{s})\rangle$ into  Eq.~(\ref{siegert}), it follows that
\begin{align}\label{1dcorrfunc}
 G^{(1)}(\bm{q}_{s},\bm{q'}_{s})\approx \sqrt{\langle \delta I(\bm{q}_{s})\delta I(\bm{q'}_{s})\rangle}.
\end{align}
Thus, $G^{(1)}(\bm{q}_{s},\bm{q'}_{s})$ can be experimentally measured by studying the correlations between the signal field's intensity fluctuations in the far-field.
\begin{figure*}[t!]
\centering
	\includegraphics[width=0.96\textwidth]{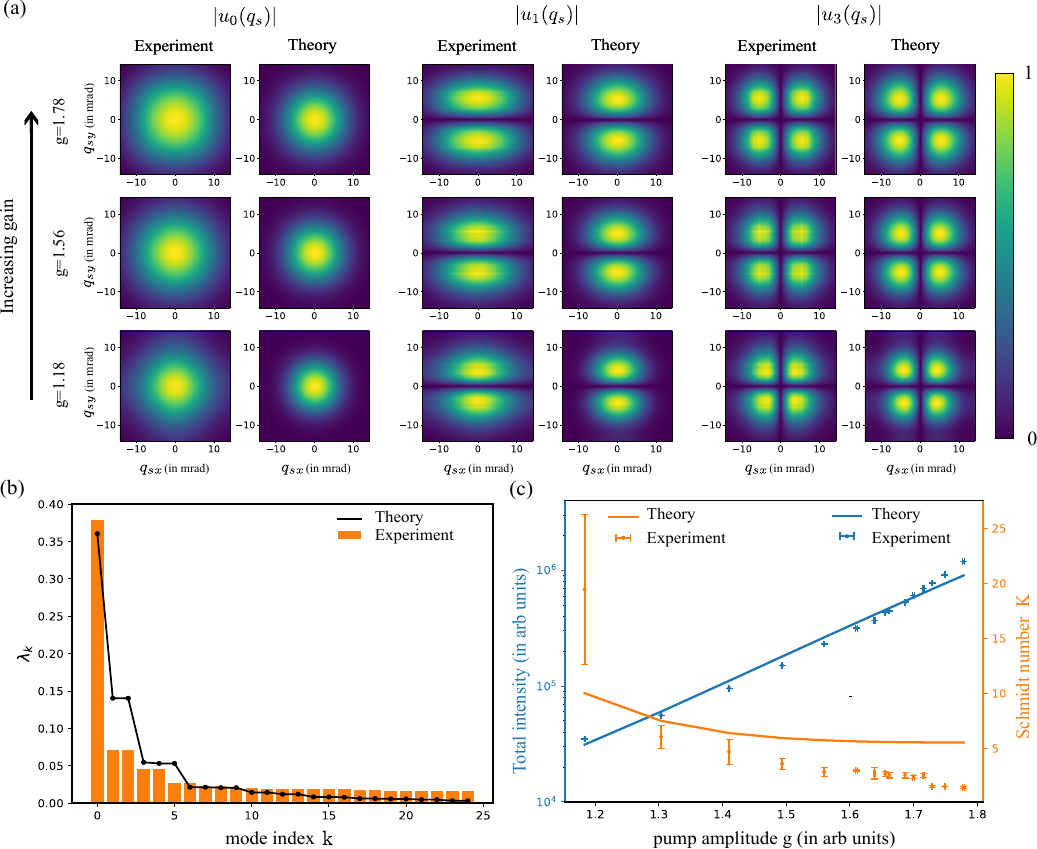}
	\caption{(a) Experimental and theoretical results depicting the spatial magnitude profiles of three Schmidt modes $\bm{u}_{0}(\bm{q}_{s}), \bm{u}_{1}(\bm{q}_{s})$, and $\bm{u}_{3}(\bm{q}_{s})$ for increasing gain values, (b) Schmidt spectrum for the gain $g=1.49$, (c) Variation of the total signal intensity and Schmidt number $K\equiv1/\sum_{k}\lambda^2_{k}$, where $\sum_{k}\lambda_{k}=1$, with respect to increasing gain.}
    \label{fig4}
\end{figure*}

In our experimental setup depicted in Fig.~\ref{fig3}(a), a 355-nm 30-ps pulsed EKSPLA PL2231 Nd:YAG laser  emits vertically-polarized pulses of light at a repetition rate of 50 Hz. These pulses are spatially filtered to obtain the desired rotationally-symmetric Gaussian profiles. The amplitude of each pulse is controlled using a combination of half-wave plate (HWP) and polarizing beam-splitter (PBS). This amplitude is measured upto an overall scaling factor using a Coherent EnergyMax USB-J-10MB-HE energy meter. The beam-waist size $w_{p}$, defined as $1/e^{2}$ width, of the Gaussian pump was measured using a Gentec Beamage-3.0 beam profiler and was found to be $w_{p}=185\,\,\mu$m at the waist plane. The pulses are incident onto a 3-mm $\beta$-barium borate (BBO) crystal cut for generating horizontally-polarized signal-idler photons for perpendicular incidence of the pump. The residual pump after the crystal is removed by means of two dichroic mirrors and dumped into beam dumps (BDs). The signal field is filtered using a filter of bandwidth 10 nm centered at 700 nm and then imaged using an Andor Ixon-897 electron multiplying CCD camera (EMCCD) at resolution 256 $\times$ 256 pixels with individual pixel size 16 $\mu$m $\times$ 16 $\mu$m. The camera was triggered using an electrical signal from the laser to ensure that each image corresponds to exactly one pulse. As depicted in Fig.~\ref{fig3}(b), we acquired 5 sets of 1000 single-shot images each of the signal far-field for each pump amplitude value. Owing to our assumption of rotational invariance, we only need to evaluate $G^{(1)}(q_{sx},q'_{sx})$ along a single diametric x-axis slice, but as each diametric slice is statistically identical, we can effectively utilize slices of all possible orientations. For a given slice, $G^{(1)}(q_{sx},q'_{sx})$ can be computed as
\begin{align}\notag
 G^{(1)}(q_{sx},q'_{sx})&\approx \frac{1}{M}\sum_{j=1}^{M}\sqrt{\langle \delta I_{j}(q_{sx})\delta I_{j}(q'_{sx})\rangle}\\\label{coinc}&\hspace{5mm}-\frac{1}{M}\sum_{j=1}^{M}\sqrt{\langle \delta I_{j}(q_{sx})\delta I_{j+1}(q'_{sx})\rangle},
\end{align}
where $M$ is the number of images in the ensemble and $\delta I_{j}(q_{sx})=I_{j}(q_{sx})-\langle I_{j}(q_{sx})\rangle$ is the intensity fluctuation of the pixel at distance $q_{sx}$ from the center in the $j$'th image and $\delta I_{M+1}(q_{sx})\equiv \delta I_{1}(q_{sx})$. Note that the above equation (\ref{coinc}) is markedly distinct from Eq.~(\ref{1dcorrfunc}) due to the presence of the second term. This term quantifies the accidental correlations that are introduced between pixels from the $j$'th image and the $(j+1)$'th image even though the two images were taken for two different pump pulses \cite{defienne2018prl,reichert2018scirep}. Consequently, this term is subtracted from the total correlations quantified by the first term to equal the true correlations $G^{(1)}(q_{sx},q'_{sx})$ on the left hand side. Incidentally, this background subtraction step also ensures that $G^{(1)}(q_{sx},q'_{sx})$ tends to zero when the separation $|q_{sx}-q'_{sx}|$ is sufficiently large. Thereafter, $G^{(1)}(q_{sx},q'_{sx})$ is diagonalized and the resulting eigenvalues and eigenmodes are tensor-multiplied as previously described to yield the full Schmidt spectrum and full 2D Schmidt modes.

In Fig.~\ref{fig4}, we present our experimental results and their comparison with theoretical predictions for different gain values. We used the parameter $C_{1}$ from Eq.~(\ref{hgpdccorrfunc}) as a fit parameter in our simulations. In addition, the scaling factor $C_{2}$ from Eq.~(\ref{gammaexp}) was chosen such that $g<1$ and $g>1$ correspond to the low and high gain regimes, respectively. The condition for collinear emission $k_{jz}=|\bm{k}_{j}|$ for $j=p,s,i$ yields $\theta^{\rm (coll)}_{p}=32.753^{\circ}$ and we chose $\theta_{p}=32.7^{\circ}$ in our simulations. The small offset is allowed because $\theta_{p}$ is approximately set by hand in experiment. In Fig.~\ref{fig4}(a), we depict the magnitude profiles of three Schmidt modes $\bm{u}_{0}(\bm{q}_{s}), \bm{u}_{1}(\bm{q}_{s})$, and $\bm{u}_{3}(\bm{q}_{s})$ for three representative gain values, namely, $g=1.18, 1.56$ and $1.78$.  We note that the Schmidt modes expand slightly with increasing gain. For instance, the experimentally measured full width at half-maximum (FWHM) of the first mode $\bm{u}_{0}(\bm{q}_{s})$ grows from $20.99$ mrad at $g=1.18$ to $26.63$ mrad at $g=1.78$. In particular, this corresponds to a $26.87 \%$ increase in the experimentally measured FWHM, and the theoretically predicted FWHM shows an increase of $24.61 \%$ in the same gain range. In Fig.~\ref{fig4}(b), we depict the Schmidt spectrum $\left\{\lambda_{k}\right\}$ over 25 modes for $g=1.49$, where we have normalized $\lambda_{k}$ such that $\sum_{k}\lambda_{k}=1$. In Fig.~\ref{fig4}(c), we show the variation of the total signal intensity and the Schmidt number $K\equiv 1/\sum_{k}\lambda^2_{k}$, which quantifies the effective width of the Schmidt spectrum. The signal intensity on the left y-axis grows exponentially with pump amplitude as expected in the high-gain regime, and the Schmidt number $K$ on the right y-axis decreases with increasing gain, which indicates a gradual reduction in the degree of spatial entanglement between the signal and idler fields with increasing gain. In all our results, we find good agreement between experimental measurements and theoretical predictions. Nevertheless, the slight mismatch between the two may be because of the fact that our theoretical calculations assume a monochromatic signal field at $700$ nm, whereas in experiment, we use a spectral filter centered at the same wavelength with a FWHM bandwidth of $10$ nm.

\section{Summary and Outlook}\label{conclusions}
In this work, we propose and demonstrate a highly efficient method for theoretically and experimentally characterizing the spatial Schmidt modes and Schmidt spectrum of the bright squeezed vacuum field produced from high-gain SPDC. The existing method adopted in Ref.~\cite{averchenko2020pra} is fully general but is computationally highly cumbersome and scales poorly with the computational input size specified by the number of pixels $N\times N$ and the number of images $M$. We restrict our attention on the specific but commonly encountered scenario where the pump field is rotationally symmetric and, consequently, the generated signal field is quasi-homogeneous and isotropic, i.e, the degree of first-order spatial coherence is translationally and rotationally invariant. In this scenario, we show that the associated symmetries can be used to significantly speed up the numerical computations involved in both the experimental and theoretical characterization procedures. In comparison to the method adopted in Ref.~\cite{averchenko2020pra}, our approach already achieves a computational speedup of over two orders of magnitude for the computational input size of $N\sim256$ and $M\sim5000$ considered in this paper, and this speedup is expected to be even more dramatic for larger input sizes. As a direct consequence of this speedup, we are able to perform the characterization efficiently and accurately for multiple different pump amplitudes and study the behavior of the Schmidt spectrum and Schmidt modes as a function of increasing gain. We find that our experimentally measured modes and spectrum are in good agreement with their theoretically predicted counterparts. Our results also clearly reveal the slight expansion of the Schmidt modes and gradual narrowing of the Schmidt spectrum as the gain increases.

In essence, our method sacrifices complete generality for a significant computational speedup in specific but commonly encountered experimental scenarios. For such scenarios, our method has the potential to be the standard choice in the future for characterizing the Schmidt modes of entangled light fields or even the coherent modes of classical light fields \cite{starikov1982josa, mandel1995cup}. In addition, our work can have important implications in the field of quantum imaging \cite{moreau2019natrevphys}. For instance, quantum imaging schemes such as sub-shot noise imaging \cite{brida2010natphot,ruo2020apl}, ghost imaging \cite{pittman1995pra,moreau2018opex}, and imaging with undetected photons \cite{lemos2014nature,fuenzalida2022quantum} face the need to improve the resolution and contrast while also preserving any genuine quantum advantage that may be present. Our work can be seen as a step towards meeting this need because it provides a quantitative characterization of the spatial correlations of bright multiphoton entangled states from high-gain SPDC and their dependence on the amplitude of the pump field. We hope that this work may eventually inform the development of novel quantum imaging schemes that could potentially beat the best available classical counterparts.
\section*{Acknowledgment}
We are thankful for discussions with Jeremy Rioux, Cheng Li, Jonathan Leach, Boris Braverman, and Maria Chekhova. We acknowledge support through the Natural Sciences and Engineering Research Council of Canada, the Canada Research Chairs program, and the Canada First Research Excellence Fund award on Transformative Quantum Technologies.  In addition, RWB acknowledges support through US National Science Foundation Award 2138174 and US Department of Energy award FWP 76295.

\bibliography{spatialschmidt}
\end{document}